\documentclass[11pt]{article}
\usepackage{graphicx}
\usepackage{latexsym}
\usepackage{amsmath}
\usepackage{amssymb}
\begin{document}

 \title{Dark Matter Annihilation Induced Gamma Ray Emission from Galaxy Cluster 1E0657-56}

 \author{C. Zhang\thanks{Email: zhangchao@mail.ihep.ac.cn}, G.-C. Liu  \\
 \small\it{College of Science, China Three Gorges University, Yichang 443002}}
\date{}
 \maketitle

\begin{center}
\begin{minipage}{130mm}
\vskip -5mm
 {\large \bf Abstract:} \, Based on minimal supersymmetric standard model, neutralino dark matter annihilation
induced gamma ray emission from galaxy cluster 1E0657-56 is calculated. The merge of bullet-like subcluster with
the main cluster is also investigated. 
     \\
\end{minipage}
\end{center}

\section{Introduction}
\par
Both astronomical and cosmological observations have concluded that Dark Matter(DM) contributes
 a quarter of the energy partition in universe and DM dominates the matter composition in galaxiy.
Among various theoretical DM particles candidates, neutralino predicted by SUSY models is the most
attractive one. Though there is no electromagnetic radiations, as the antipaticle of itself, 
they can annihilate into a pair of photons or other particles which eventually decay to the final state
photons. Theoretically, the corresponding gamma ray flux is proportional to the square of the dark
matter density. Therefore, such a signal can most probably be observed from the center of the galaxies
or substructure in galaxies.
\par
Galaxy cluster 1E0657-56 is a very massive cluster who has an exceptional bullet-like subcluster in it.
The weak lensing observation of bullet cluster indicates the separation of baryonic mass component 
and unseen matter induced by the merger, which provided a direct empirical proof of the existence
of DM\cite{proof}. This encourage us to study the corresponding DM signals from it. 
In this work, we calculate the DM annihilation induced $\gamma$-ray flux from 
the central direction of main cluster and bullet cluster, respectively.
\section{Gamma Ray Flux from Dark Matter Annihilation}
\par
Dark matter particle is usuallay considered as the Light Supersymmetric Particle(LSP) which,
in most supersymmetric scenarios, is regarded as the neutralino $\chi$. Theoretically, gamma
ray from $\chi\chi$ annihilation is expected to exhibit both monochromatic and continuum
spectrum while the later part makes the dominant contribution and therefore becomes the major concern
of this work. Suppose that the dark matter is concentrated in a spherical halo of virial radius
$r_{vir}$ and denote the density profile $\rho_{\chi}$ as a function of the radial distance, then
energy spectrum of gamma ray from DM annihilation in a galaxy can be written as\cite{anni}
\begin{eqnarray}
\frac{\mathrm{d}\phi_{\gamma}(E,\psi)}{\mathrm{d}E_{\gamma}} &=& 
      \sum_{F}\frac{\mathrm{d}N}{\mathrm{d}E}b_{F}
      \frac{\langle \sigma\upsilon\rangle}{2m^{2}_{\chi}}
	\frac{1}{4\pi d^{2}}\int_{\Delta\Omega}d\Omega\int_{l.o.s}\rho_{\chi}^{2}(r)r^{2}dr\nonumber\\
    &=& 1.87\times 10^{-11}\frac{N_{\gamma}(E)\langle \sigma\upsilon\rangle}{10^{-29}cm^{3}s^{-1}}	
	\left(\frac{10GeV}{m_{\chi}}\right)^{2} \nonumber\\
    && \cdot \frac{1}{8.5 kpc}\left(\frac{1}{0.3 GeV cm^{-3}}\right)^{2}
        \frac{1}{4\pi d^{2}}\int_{\Delta\Omega}d\Omega\int_{l.o.s}\rho_{\chi}^{2}(r)r^{2}dr	
	, \label{susy}
\end{eqnarray}
where $N_{\gamma}(E)=\sum_{F}\frac{\mathrm{d}N}{\mathrm{d}E}b_{F}$ describes the differential energy spectrum
of the gamma ray from DM annihilation for all final states $F$, $d$ refers to the distance to the galactic center.
$\Delta\Omega$ represents the solid angle for a given angular resolution of the detector. 
Actually, Eq.(\ref{susy})
can be divided by two parts, one called ``particle factor" which is exclusively determined by the feature
of particle, and the other called ``cosmological factor" which decided by the density distribution of
the local dark matter halo. This allow us to calculate them separately. 
\par
In the Minimal Supersymmetric Standard Model(MSSM),
certain SUSY parameter space gives the information of the mass and the interaction strength of DM particles.
Generally, SUSY parameter space is decided by seven independent SUSY parameters 
$\mu, M_{2}, M_{A}, m_{\tilde{q}}, \tan\beta, A_{t}$ and $A_{b}$. 
For the particle factor part, we perform a random
scanning of these parameters within the corresponding ranges by using the package DarkSUSY\cite{darksusy}.
The value of these parameters are constrained by the theoretical consistency. Furthermore, the result is 
checked by current accelerator data and experimental observations. One of strong limits comes from the 
observational data of the relic abundance of the cold dark matter, 
$\Omega_{CDM}h^{2}=0.113^{+0.008}_{-0.009}$,  
specifically\cite{abundance}. We make use of $3\sigma$ error bar range to limit the parameter space,
 i.e., $0.086<\Omega_{\chi}h^{2}<0.137$. 
The relic bundance of neutralino particle less than the minimal limit indicates a subdominant DM component.
Then, a rescaling of DM density will be done as $\rho(r)\rightarrow\xi\rho(r)$ with 
$\xi=\Omega h^{2}/(\Omega_{\chi}h^{2})_{min}$\cite{Bi}.
\section{Gamma Ray Emission from Galaxy Cluster 1E0657-56}
\par
The main cluster of galaxy cluster 1E0657-56 
is found that the halo density contribution is well fitted by
NFW\cite{nfw} profile with concentration $c_{200}=3.0$ and virial radius $r_{200}=1.64h^{-1}$Mpc, corresponding 
to a virial mass and velocity of $M_{200}=1.5\times 10^{15}M_{\odot}$ and
 $v_{200}=2380 kms^{-1}$\cite{clowe04}. The bullet substructure, located about $0.48h^{-1}$ Mpc to the 
centre of the main cluster,
is the most massive substructure in the cluster with the $M_{200}=1.5\times 10^{14}M_{\odot}$, $c_{200}=7.2$
and $r_{200}=0.78h^{-1}$Mpc.
 The NFW density profile can be described as 
\begin{equation} \label{nfwprofile}
   \rho_{\chi}(r)=\frac{\rho_{s}}{(r/r_{s})(1+r/r_{s})^{2}},
\end{equation}     
where $\rho_{s}$ and $r_{s}$ are the scale density and scale radius respectively. These two free parameters
can be determined by the measurements of the virial mass and the concentration parameter of the halo.
In order to avoid the unphysical singularity when $r\rightarrow 0$,
 we take a core radius $r_{min}=10^{-11}$ kpc, within which the DM density keep invariant\cite{core}.
\par
Cluster 1E0657-56 is a very massive cluster formed by a plenty of sub-clusters.
Dark matter halo clustered in small length scales was wildly discussed\cite{sub1,sub2,sub3,sub4}. 
Since clumpy sub-halo will greatly enhance the $\gamma$-ray emission, the effect including substructure
should be considered. The clumpy magnitude of dark halo is mainly determined by the product of two parameters, 
$f$ and $\delta$. where $f$ is defined as the fraction of clumpy component in the total dark halo, and 
$\delta$ is a dimensionless parameter which gives the effective contrast between the sub-halo
density and the local halo density $\rho_{0}$, $\delta=\rho_{cl}/\rho_{0}\sim O(1000)$.
The factor $f$ is estimated to be around between $5\%\sim 20\%$ in the literature\cite{clumpy1,clumpy2,clumpy3}.
Following the approach of Ref.\cite{sub3} and the references therein, the cosmological factor
$\langle J(\psi)\rangle_{\Delta\Omega} $ can be rewrite as 
\begin{equation}
\langle J(\psi)\rangle_{\Delta\Omega}
       = \frac{1}{8.5 kpc}\left(\frac{1}{0.3 GeV cm^{-3}}\right)^{2}
        \frac{\rho_{0}f\delta}{4\pi d^{2}}
        \int_{\Delta\Omega}d\Omega\int_{l.o.s}\rho_{\chi}(r)r^{2}dr, \label{susy2}
\end{equation}
where we take the mass fraction of sub-structure as $f_{main}=10\%$ and $f_{bullet}=50\%$\cite{nature}
 in our calculation.
\section{Results and Discussions }
\begin{figure}[htp]
\includegraphics[width = \textwidth]{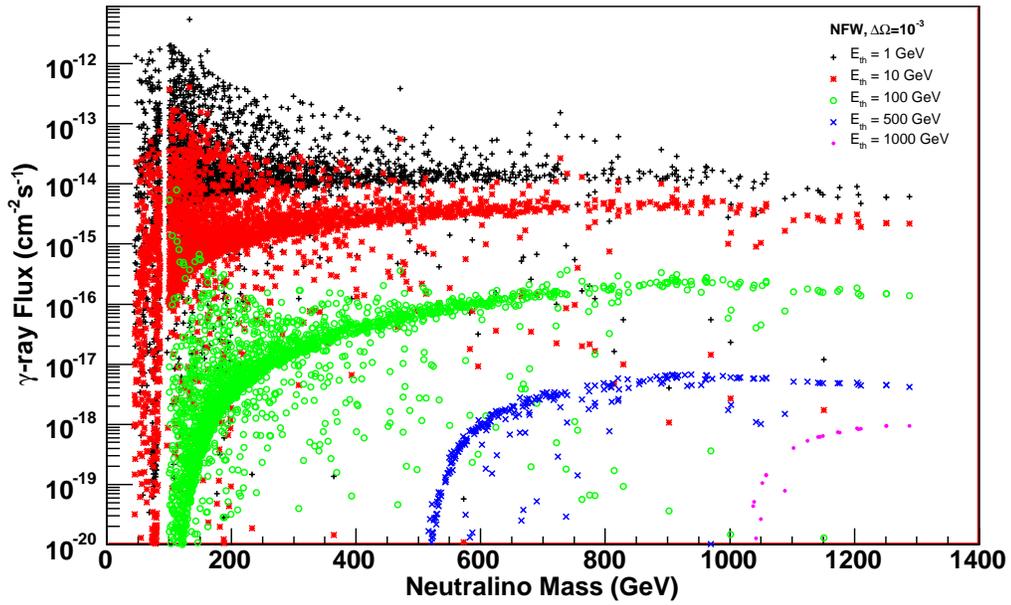}
\caption{\label{result}  The integrated gamma ray flux from the centre of galaxy cluster 1E0657-56
above the threshold energy of 1,10,100,500,1000GeV, where NFW halo profile and the angular
resolution of $\Delta\Omega = 10^{-3}$ is set. Each neutralino mass refers to a set of 
SUSY parameter space which checked by all the theoretical and experimental limit.}
\end{figure}

Based on the theoretical frame given above, dark matter induced $\gamma$-ray flux 
from centre of main cluster and bullet subcluster are calculated, respectively.
The central distance between the two cluster is about $0.48 h^{-1}$ Mpc which
is equivalent to the angle of $6.322\times 10^{-4} $ between the light of sight.
Furthermore, the overlap of the two cluster is considered where 
the angular resolution of $\Delta\Omega= 10^{-3}$ is applied.
For the main cluster, we get 
$\langle J(\psi)\rangle_{\Delta\Omega} = (3.61\times 10^{-3})_{clumpy} + (6.56\times 10^{-4})_{overlap} $.
While we get 
$\langle J(\psi)\rangle_{\Delta\Omega} = (1.56\times 10^{-3})_{clumpy} + (2.79\times 10^{-3})_{overlap} $
 for the bullet subcluster. 
This shows that the two have almost the same intesity of gamma ray emission. We consider the constrains 
come from both particle physics and astrophysics and the final result 
 is presented in Figure \ref{result}.
\par
The result shows that, for the neutralino annihilation induced $\gamma$-ray emission, 
galaxy cluster 1E0657-56 has a double peak structure on the spacial distribution.  
For those Imaging Air Cherenkov Telescopes(IACTs), their sensitivity can achieve 
$\sim 10^{-12}cm^{-2}s^{-1}$ in the $100 GeV\sim 10 TeV$ region. Even though, there are 
still several orders of magnitude difference to be detectable, which is hard to reach 
simply by increasing statistical time.
\par
In comparison with cherenkov technique, EAS experiments such as Milagro\cite{milagro}
and ARGO\cite{argo} have the advantage in large field of view and higher duty cycle
which enable them to posses the capability in observing simultaneously a large number of 
sources of sub-structures. Therefore, they have the potential in improving the detection
sensitivity of gamma ray from DM annihilation.  
\section*{Acknowledgments}
 This work is supported in part by the NSF of China under the grant No.10747142.

\end{document}